\documentclass[10pt,prx,twocolumn,eqsecnum,amssymb,amsmath,showpacs,a4paper,superscriptaddress,nofootinbib,floatfix]{revtex4-2}
\usepackage{graphicx}
\usepackage{amsfonts}
\usepackage{amsmath}
\usepackage[linktocpage=true]{hyperref} 
\usepackage{units}
\usepackage{color}
\usepackage{dcolumn}
\usepackage{bm}
\usepackage{cleveref}
\usepackage{braket}
\usepackage{tikz}
\crefname{section}{§}{§§}
\Crefname{section}{§}{§§}
\usepackage{mathtools}
\usepackage{chngcntr}
\counterwithout{equation}{section} 
\usepackage[normalem]{ulem}
\usepackage{ulem}
\usepackage[version=4]{mhchem}
\usepackage[percent]{overpic}
\usepackage{tabularx}
\usepackage[most]{tcolorbox}
\usepackage[utf8]{inputenc}
\usepackage{natbib}
\usepackage{pdfpages}
\usepackage{pgffor}
\usepackage{etoolbox} 

\makeatletter
\patchcmd{\@outputpage@head}{\@ifx{\LS@rot\@undefined}{}{\LS@rot}}{}{}{}
\makeatother

\usepackage{xcolor}
\hypersetup{  colorlinks,
              linktoc         = page, 
              urlcolor        = {green!80!blue},
              linkcolor       = {orange!60!black}, 
              urlcolor        = {cyan!80!black}, 
              citecolor       = {cyan!80!black}, 
              anchorcolor     = {yellow}
}

\usepackage{lineno}

\usepackage{bm}
\usepackage{comment}

\begin{document}

\title{Multiplexed digital holography for fluid surface profilometry}

\author{August Geelmuyden}
\affiliation{School of Mathematical Sciences, University of Nottingham, University Park, Nottingham, NG7 2RD, UK}
\author{Vitor S. Barroso}
\affiliation{School of Mathematical Sciences, University of Nottingham, University Park, Nottingham, NG7 2RD, UK}
\author{Sreelekshmi C. Ajithkumar}
\affiliation{School of Physics \& Astronomy, University of Nottingham, University Park, Nottingham, NG7 2RD, UK}
\author{Anthony J. Kent}
\affiliation{School of Physics \& Astronomy, University of Nottingham, University Park, Nottingham, NG7 2RD, UK}
\author{Silke Weinfurtner}
\affiliation{School of Mathematical Sciences, University of Nottingham, University Park, Nottingham, NG7 2RD, UK}
\affiliation{Centre for the Mathematics and Theoretical Physics 
of Quantum Non-Equilibrium Systems, 
University of Nottingham,
Nottingham NG7 2RD, 
UK}

\begin{abstract}
Digital holography (DH) has been widely used for imaging and characterization of micro and nanostructures in materials science and biology and has the potential to provide high-resolution, non-destructive measurements of fluid surfaces as well. Digital holographic setups capture the complex wavefronts of light scattered by an object or reflected from a surface, allowing for quantitative measurements of their shape and deformation. However, their use in fluid profilometry is scarce and has not been explored in much depth. We present an alternative usage for a DH setup that can measure and monitor the surface of fluid samples. Based on DH reflectometry, our modelling shows that multiple reflections from the sample and the reference interfere and generate multiple holograms of the sample, resulting in a multiplexed image of the wavefront. The individual interferograms can be isolated in the spatial-frequency domain, and the fluid surface can be digitally reconstructed from them. We further show that this setup can be used to track changes in the surface of a fluid over time, such as during the formation and propagation of waves or evaporation of surface layers.
\end{abstract}

\maketitle

\section{Introduction}

The need for fast and accurate measurements of surfaces has exploded with the advent of automatic production lines and progress in computer vision. Today, popular methods are based on correlating images taken from slightly displaced vantage points~\cite{sutton2009image}, local deformations of periodic patterns~\cite{takeda1982fourier,su2001fourier}, and time-of-flight imaging. Despite such great advances in profilometry, precision detection of fluid interfaces, usually between liquid and gas, remains a challenge and has resulted in various developments in experimental fluid studies. Some notable methods applied to fluids include Fourier Transform Profilometry (FTP) and Digital Image Correlation (DIC). For instance, the former has been implemented using the projection of fringe patterns~\cite{cobelli2009global} or a background reference checkerboard~\cite{Wildeman2018}. While image correlations using random backdrops have also been used to reconstruct fluid surfaces~\cite{2009Moisy}. 

Over the last two decades, an interesting avenue for high-precision profilometry has appeared in the field of digital holography~\cite{kimPrinciplesTechniquesDigital2010,dicaprioHolographicImagingUnlabelled2014,kreisApplicationDigitalHolography2016,taharaDigitalHolographyIts2018,paturzoDigitalHolographyMetrological2018,nehmetallahLatestAdvancesSingle2020,zhangReviewCommonpathOffaxis2021,Zeng:21}. Holography, the complete reconstruction of the optical wavefront using diffraction theory, was introduced by Gabor in 1948 \cite{gaborNewMicroscopicPrinciple1948}. In the years that followed, Gabor demonstrated the ability to extract 3D information from a 2D hologram to regain focus and spatial resolution \cite{gabor1949microscopy,Gabor_1951}. The first successful image reconstruction using digital holography appeared already in 1967 \cite{goodmanDigitalImageFormation1967}. However, it was not until 1994 that Schnars and Jüptner introduced the first ever digitally reconstructed hologram using a CCD camera \cite{schnarsDirectRecordingHolograms1994}, based on the off-axis method proposed by Leith~\cite{leithReconstructedWavefrontsCommunication1962}. Subsequently, Cuche et al. introduced the possibility of using digital holography (DH) as a quantitative phase measurement \cite{cucheDigitalHolographyQuantitative1999}, now called off-axis digital holography. Spatially modulated phase-fronts are often used in the field, sometimes carrying several modulations that can be separated in the spatial-frequency domain (SFD)\cite{yuanResolutionImprovementDigital2011,shaFastReconstructionOffaxis2014,micoSpatiallymultiplexedInterferometricMicroscopy2014,rubinSixpackOffaxisHolography2017,dardikmanMultiplexedOffaxisHolography2019}.

Here, we extend fluid profilometry and digital holography methods to develop a detection scheme for fluid surfaces. When one arm in a Michelson interferometry setup probes a fluid sample, the resulting hologram is a multiplexed composition of several interference patterns, whose phases are proportional to the profile of the fluid surface. We present a systematic method to harvest the latter from modulations on the carrier peaks of each pattern, which can be singled out in the spatial frequency domain. By employing relative off-axis adjustments of the beam paths, we can optimise the separation of the carriers in the frequency domain. The relative demodulation between consecutive time frames can then be used to accurately reconstruct variations on the fluid surface.

\section{Modelling and Simulation}\label{sec:setupModel}

\begin{figure}
    \centering
    \includegraphics[width= \linewidth]{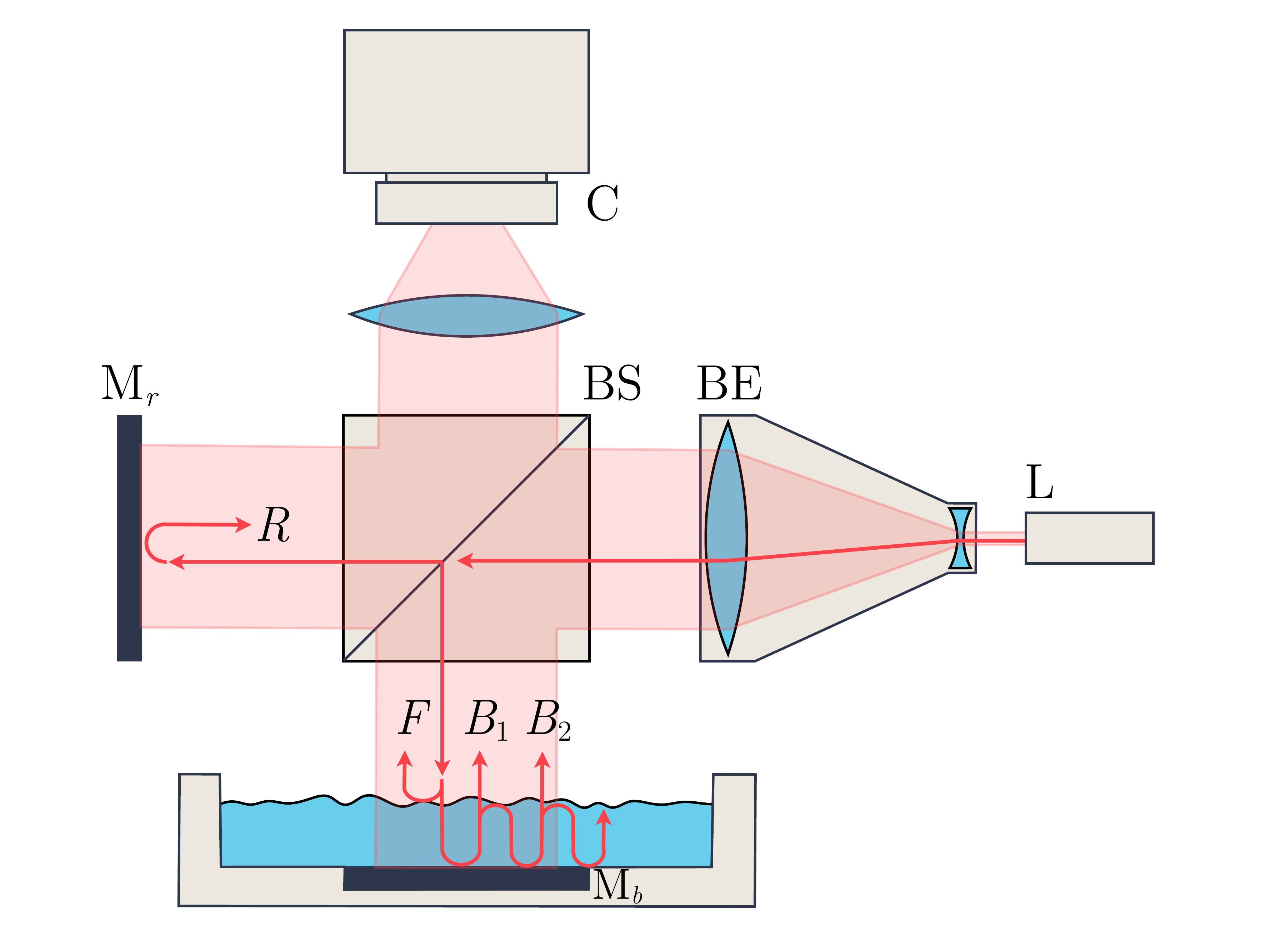}
    \caption{Schematic of proposed setup. A laser source (L) emits a collimated beam that expands to the desired diameter by passing through a Beam Expander (BE). A Beam Splitter (BS) divides the beam into two, namely a probe going through the fluid sample and a reference ($R$), both reflecting from adjustable mirrors (M). The reflected trajectories recombine in the beam splitter and the resulting beam is captured by a camera (C). In the figure, the multiple beams reflected from the fluid surface ($F$) and the probe mirror ($B_1$, $B_2$) contribute to the multiple interference peaks discussed in section 2\ref{multipeaks}.}
    \label{fig:holo:schematic}
\end{figure}

As commonly employed in off-axis Digital Holography (DH), we begin with a Michelson interferometer, shown in Fig.~\ref{fig:holo:schematic}. In this setup, a laser beam of wavelength $\lambda$ is expanded to the desired width before being parted into two arms at the beam splitter. One acts as a reference ($R$) by reflecting from a mirror (M$_r$). The other arm probes the sample, where the beam is reflected partially by the fluid surface and partially by the mirror (M$_b$). The general strategy of off-axis DH that we explore here is to slightly misalign the reference with respect to the probe beam resulting in a plurality of linear interference patterns. Hence, after the two arms recombine in the beam-splitter, a digital camera (C) captures the interferograms (or holograms). In what follows, we briefly discuss the relations between the sample's height profile and the fringe patterns resulting from the path differences of various components of the probe beam. 
\subsection{Phase differences and sample's height profile \label{multipeaks}}

We begin by examining the optical paths of different types of rays as they traverse an air-fluid interface $z=h(t,\mathbf{r})$. More precisely, consider a ray that starts at some reference height $z=H$ with transverse horizontal coordinate $\mathbf{r}$, and moves downwards to the fluid interface $z=h$, where it will be partially transmitted, partially reflected. We shall refer to the latter as an $F$-type ray. The transmitted ray travels through the fluid and reaches a submerged plane mirror $M_b$ at the bottom of the basin. Consequently, the fluid interface and the mirror act as a cavity, producing a collection of partially transmitted rays $B_j$, labelled by the number $j$ of internal reflections inside the fluid (see Fig.~\ref{fig:holo:schematic}). At the beam splitter, the probe arm, containing $F$ and $B_j$ rays, recombines with the reference arm $R$ reflected from the mirror $M_r$, which yields a beam exhibiting four families of interferograms, namely: $RF$, $RB_j$, $FB_j$, $B_\ell B_j$. For any pair $AB$ of these rays, their phase difference $\Phi_{AB} = \Phi_A-\Phi_B$ is proportional to their optical path differences. Camera C captures the interfered beam and generates an image $\bm{Y}=(Y_{ij})$ displaying the combined intensity profile of all interferograms, for positive integer pixel locations $(i,j)$. We shall assume all phase differences $\Phi_{AB}$ to have small modulations $\phi^{(AB)}(t,\mathrm{r})$ around a stationary planar phase $\mathbf{k}_{AB}\cdot \mathbf{r}$. That is, $\Phi_{AB}(t,\mathbf{r}) = \phi^{(AB)}(t,\mathbf{r}) + \mathbf{k}_{AB}\cdot\mathbf{r}$, so that
\begin{equation}
    Y_{ij} 
    =Y_0\left(1+
    \sum_{\substack{AB \\ A\neq B}} \mathcal{A}_{AB}\exp\left(i\mathbf{k}_{AB}\cdot \mathbf{r}_{ij}+i\phi_{ij}^{(AB)}\right) \right)
    + \delta Y_{ij} 
    ,\label{eq:holo:image}
\end{equation}
where $\delta Y_{ij}$ is a stochastic variable that describes the noise in the image, for an arbitrary intensity $Y_0$ and amplitudes $\mathcal{A}_{AB}$ (see Supplemental Document for their formulae). 

Let us now consider that both mirrors are slightly tilted with respect to the optical axis $\mathbf{\hat{z}}$, i.e., their normal unit vectors $\mathbf{m}_b$ and $\mathbf{m}_r$ are such that $\mathbf{m}_b\cdot \mathbf{\hat{r}}\ll 1$ and $\mathbf{m}_r\cdot \mathbf{\hat{r}}\ll 1$.
In these conditions, one can show that for weak surface slopes, i.e., $|\nabla h|\ll 1$, the phase differences carried by each family of holograms read (cf.~\cite{ColombDHR2010,colombDHR2012})
\begin{subequations}
\label{eq:holo:interferogramsPhases}
\begin{align}
    \Phi_{RF} &= 2k_0 n_1\left( h+ \mathbf{m}_r\cdot \mathbf{r}+ \Delta H \right),
    \label{eq:phase0_RF}
    \\
    \Phi_{RB_j} &= 2k_0[(n_1-jn_2)h + (n_1\mathbf{m}_r-jn_2\mathbf{m}_b)\cdot \mathbf{r}+n_1\Delta H],
    \label{eq:phase0_RBj}
    \\
    \Phi_{FB_j} &= 2k_0 n_2 [jh + j\mathbf{m}_b\cdot \mathbf{r}],
    \label{eq:phase0_FBj}
    \\
    \Phi_{B_\ell B_j} &= 2k_0  n_2[(\ell-j)h + (\ell-j)\mathbf{m}_b \cdot\mathbf{r} ],
    \label{eq:phase0_BlBj}
\end{align}
\end{subequations}
where $\Delta H \equiv H_R-H$ is the difference in arm length of the reference beam $H_R$ and the probe beam $H$, $n_1$ and $n_2$ are the refractive indices of the surrounding medium (e.g.~air) and the fluid, respectively. We define the laser's wavenumber as $k_0\equiv 2\pi/\lambda$. 

From~\eqref{eq:holo:interferogramsPhases}, we see that the planar phases $\mathbf{k}_{AB}\cdot\mathbf{r}$ are proportional to $\mathbf{m}_b\cdot \mathbf{r}$ and $\mathbf{m}_r\cdot \mathbf{r}$. In the presence of non-vanishing small tilts in the normal vectors $\mathbf{m}_r$ and $\mathbf{m}_b$, an image $Y_{ij}$ obtained from the camera displays the collection of distinguishable fringe patterns, with phases given by~\eqref{eq:holo:interferogramsPhases}. One such simulated image frame is given in panel (a) of Fig.~\ref{fig:holo:syntheticData}. By performing a two-dimensional Fourier transform $\mathcal{F}$ of the image, i.e. $\hat{Y}_{ij}\equiv\mathcal{F}[Y]_{ij}$, one can inspect its spectrum in spatial-frequency domain (SFD), where the holograms separate into distinct spatial carrier frequencies. Fig. ~\ref{fig:holo:syntheticData} (b) shows the frequency domain of the corresponding simulated image $Y_{ij}$ in Fig.  ~\ref{fig:holo:syntheticData} (a)  . 
The signal, the interfacial height $h(t,\mathbf{r})$ here, appears as a modulation of the spatial carrier $\mathbf{k}_{AB}$, for any family of interferograms $AB$. Thus, the variations $\delta \phi_{AB}$ in the phase of a carrier due to deformations $\delta h$ on the sample's interface are given by
\begin{align}
    &\left(\frac{\delta\phi_{AB}}{2\pi}\right) = 2\alpha_{AB}\left(\frac{\delta h}{\lambda}\right),\label{eq:holo:phasetoheight}\\
    &\alpha_{AB}= 
    \begin{cases}n_1, & \text { for } AB=R F; \\ 
    j~n_2, & \text { for } AB=F B_j; \\
    n_1-j~n_2, & \text { for } AB=R B_j.\end{cases}\nonumber
\end{align}
When there are multiple rays with different carrier frequencies, as in \eqref{eq:holo:interferogramsPhases}, the resulting intensity is referred to as Multiplexed Off-axis Digital Holography (see, e.g. \cite{rubinSixpackOffaxisHolography2017}). Here multiplexing refers to the fact that several independent measurements can be packed into a single snapshot of the intensity due to their separation in SFD. While the multitude of rays resulting from partial reflection has been exploited in existing digital holography techniques, called Digital Holographic Reflectometry~\cite{ColombDHR2010,colombDHR2012}, there is one important difference from the approach presented here. By introducing a tilt $\mathbf{m}_b$ of the submerged mirror, the collection of partially reflected rays separate and isolated off-axis holograms may be formed by any pair of rays. With the adjustability of the submerged mirror, we may use a simple Michelson interferometer in Fig. \ref{fig:holo:schematic} to perform simultaneous, independent measurements of the surface by virtue of the multiplexing in \eqref{eq:holo:interferogramsPhases}. That is, our method is a multiplexed off-axis extension of digital holographic reflectometry. It is worth noting that because of the tilts, different holograms will image slightly shifted regions of the fluid interface. Hence, the reconstructed profile $h_{AB}$ of an interferogram $AB$ is related to the fluid surface profile $h$ by $h_{AB}(t,\mathbf{r})=h(t,\mathbf{r}+\bm{\delta}_{AB})$, where $\bm{\delta}_{AB}$ is a spatial displacement related to the physical tilt between rays $A$ and $B$ (see Fig.~\ref{fig:holo:syntheticData}). It is worth noting that a similar argument would apply to variations in time. However, the time that the $B_j$ rays take to travel through the fluid, reflect from the bottom mirror or the surface and recombine with other rays is negligible compared to the time scales of low-frequency waves on the fluid surface. 

\subsection{Numerical phase-tracing and simulated holograms }


\begin{figure}[t!]
\includegraphics[width= 1\linewidth]{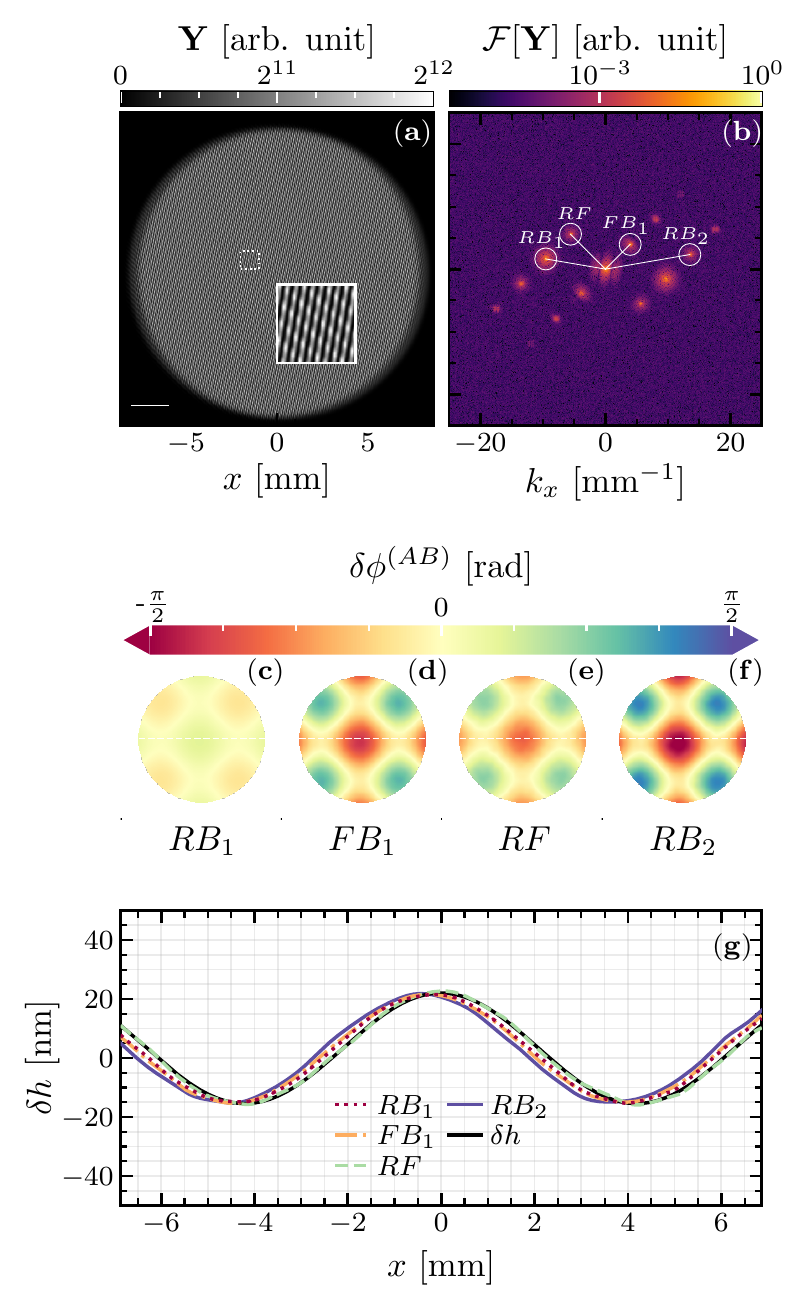}
\caption{ \textbf{Numerical simulation.}   
    In \textbf{(a)}, we display a numerically simulated intensity image $\bm{Y}$ with an inset highlighting an expanded region (white dashed rectangle) of the image. A white line on the lower-left corner depicts a length of $2$ mm.
    The spatial Fourier transform $\bm{\hat{Y}}$ of the simulated intensity $\bm{Y}$ is shown in \textbf{(b)}, where the brightness of each pixel is proportional to the logarithmic amplitude $\log |\bm{\hat{Y}}|$.
    White lines are drawn to illustrate the carrier frequencies for the holograms $RB_1$, $RF$, $FB_1$ and $RB_2$ in line with~\eqref{eq:holo:interferogramsPhases}. In \textbf{(c)}-\textbf{(f)}, we show the four corresponding phase differences $\delta\phi^{(AB)}$ between two consecutive time frames for holograms $AB$ with a shared colorbar. The visibly correlated spatial dependence of the phases is proportional to the change in fluid height $\delta h(\mathbf{r})$. Slices of the phases at the white, dashed lines in \textbf{(c)}-\textbf{(f)} are shown in \textbf{(g)} already converted to height variations $\delta h$ using the prefactors obtained from the model of~\eqref{eq:holo:phasetoheight}. The black line indicates the ground-truth height change $\delta h$ used to generate the simulation. The spatial displacement between the reconstructed curves and the ground truth confirms the expected shift due to the relative tilts between the sample and reference mirrors.
    }
\label{fig:holo:syntheticData}
\end{figure}

We initially investigate the validity of the proposed model by numerically simulating the result of an experimental acquisition. This can be done by iteratively propagating a mesh of rays through a mathematical formulation of the setup presented in Fig.~\ref{fig:holo:schematic} (see Supplemental Material for details). We model the two mirrors as tilted planes $z=-\mathbf{m}_b\cdot \mathbf{r}$ and $z=-\mathbf{m}_r\cdot \mathbf{r}$, and define the interface $z=h(t,\mathbf{r})$ for some analytic function $h(t,\mathbf{r})$. 
We then start with a cartesian mesh of reference positions for the initial ray and all directions to be equal and perfectly downward-facing. Each ray in the mesh may then be propagated linearly to the intersection point $z=h(t,\mathbf{r})$, where the Fresnel boundary conditions can be used to split the initial amplitude in a reflected and refracted component. The refracted mesh of rays produces new rays at each reflection of the free surface from below. At every interface encountered, the amplitudes are modified, and at every propagation of length between interfaces, the phase attains a contribution proportional to the optical path difference.
By collecting all rays that reach the detector plane at $z=H$ within a number $N_i$ of iterations and interpolating them to shared horizontal coordinates, we are left with a collection of discretized estimates for the rays $R,F,\{B_{j}\}$. A synthetic image may then be produced by combining the intensities of all families of interferograms and sampling from a Poisson distribution to include noise in the simulation. By evolving the simulated fluid surface in time intervals of $\delta t$, we obtain consecutive frames of the intensity image $Y_{ij}$. The result of this procedure for the fluctuation in height profile $\delta h (t_0,\mathbf{r}) = h(t_0+\delta t,\mathbf{r})-h(t_0,\mathbf{r})$ for two consecutive times of standing waves in a square basin is shown in Fig.~\ref{fig:holo:syntheticData}.

\section{The Detection Method}

We turn to the reconstruction of different phase fields from a digital image. We discuss in this section the canonical approach, which in the context of digital holography is referred to as the Angular Method~\cite{kimPrinciplesTechniquesDigital2010}, and Fourier Demodulation~\cite{Takeda:83} in that of fluid profilometry. We then explore the alternatives for identifying the various families of peaks in an experimental setup.

\subsection{Digital recovery of the phase\label{sec:demodulation}} 
With the Michelson setup presented in Fig.~\ref{fig:holo:schematic}, a digital image $Y_{ij}$ as in~\eqref{eq:holo:image} can be obtained. The Fourier spectrum of $Y_{ij}$ displays intensity peaks around the $\mathbf{k}_{AB}$ vectors, whose positions are proportional to $\mathbf{m}_r$ and $\mathbf{m}_b$, as per~\eqref{eq:holo:interferogramsPhases}. For convenience, we label each peak by an integer $m$, and provided with some function $G_{m}(\mathbf{k})$ that is non-vanishing around the carrier $\mathbf{k}_{m}$ only, we define a Fourier filter $F_{m}\equiv \mathcal{F}^{-1}G_{m}(\mathbf{k}) \mathcal{F}$ around each peak. In line with the Fourier Demodulation, see e.g.~\cite{Wildeman2018,takeda1982fourier}, we select a reference image $(Y_0)_{ij}$ filtered around one of the carriers, i.e., $F_{m}[Y_0]_{ij}$, to recover the phase variations $\Delta\phi_{ij}^{(m)}$ with respect to the reference. Neglecting noise contributions, the outcome is then:
\begin{equation} \label{eq:syntheticReconstruction}
    \Delta\phi_{ij}^{(m)} 
    = \text{Im}\left\{ \log \left[ (F_{m}[Y]_{ij})(F_{m}[Y_0]_{ij})^{*} \right]\right\}. 
\end{equation}
Numerically, this equation will only yield values between $-\pi$ and $\pi$. Thus, if the change in phase $\Delta\phi_{ij}^{(m)}$ resulting from a height variation $\delta h$ in~\eqref{eq:holo:phasetoheight} is larger than one period of $2\pi$, then a phase unwrapping algorithm, such as that of \cite{herraezFastTwodimensionalPhaseunwrapping2002}, is required.

One way to avoid the issue of phase wrappings is to consider changes in phase from some reference time $t_0$ to the time $t$ in question. If the carrier's position does not drift from $t_0$ to $t$, i.e. $\mathbf{k}_m(t_0)=\mathbf{k}_m(t)$, then, the reconstructed phase difference reads
$\Delta \phi_{ij}^{(m)} \equiv \phi_{ij}^{(m)}(t)-\phi_{ij}^{(m)}(t_0)$.
Given a sequence of images $\{Y_{ij}(t_n)\}_{n=1}^{N}$ taken at equidistant times $\delta t \equiv t_{n+1}-t_{n}$, there are two canonical choices for the reference time $t_0$. First, one may choose $t_0$ to be a constant reference, e.g. the initial frame $t_0=t_1$. We shall refer to this approach as the \textit{absolute reconstruction}. Alternatively, one can choose the previous image as the reference, i.e. $t_0 = t-\delta t$. We shall refer to this approach as the \textit{relative reconstruction}. Lastly, we shall refer to the full reconstructed phase $\phi_{ij}^{(m)}$ using a synthetic reference image of a plane wave $\exp(i\mathbf{k}_m\cdot\mathbf{r})$ as \textit{synthetic reconstruction}, which is always determined up to a global phase wrapping value $\ell_m$. In the simulated data of Fig.~\ref{fig:holo:syntheticData}, we employed the relative reconstruction to obtain the phases displayed in (c)-(f) (See Supplemental Document for more details).


\subsection{Peak identification and prefactor estimates \label{peakid}}
There are two different approaches which one may use to identify the peaks and estimate the prefactors, and hence the height fluctuations $\delta h(\mathbf{r},t)$. The first one is an experimental procedure that can be followed when acquiring data and will be discussed in section 4\ref{exppeakid}.
When manipulating the experimental setup is not possible or desirable, the second method may be used by employing a statistical tool called Principal Component Analysis (PCA) ~\cite{jolliffe1990principal,jolliffe2013principal}. Along this line, the holograms may be identified directly from the reconstructed phases $\phi_{ij}^{(m)}$. From~\eqref{eq:holo:phasetoheight}, we note that the phases $\phi_{ij}^{(m)}(t)$ scale linearly with the surface height $h(t,\mathbf{r}_{ij})$. Thus, we can recover the proportionality prefactors between measured phases and the profile of the fluid surface and use them to estimate the refractive indices. The collection of reconstructed phases $\{\phi_{ij}^{(m)}\}_{m=1}^{M}$ can be seen as an $M$-dimensional vector field $\mathbf{p}=\left(\phi_{ij}^{(1)},\dots, \phi_{ij}^{(M)}\right)$. Spatial or temporal fluctuations $\delta h$ in the surface height $h$ will result in a change in $\mathbf{p}$ given by $\delta\mathbf{p} = 2k_0\mathbf{v}\delta h$, for a constant vector $\mathbf{v} = (\alpha_1,\dots, \alpha_M)$, which depends on the refractive indices exclusively. \\

We employ to this configuration a Principal Component Analysis (PCA). For that, we consider the spectrum $C\mathbf{v}_k=\lambda_k\mathbf{v}_k$ of the covariance matrix $C_{\ell m}=\left\langle X_{ij}^{(\ell)}X_{ij}^{(m)}\right\rangle$, for centralised variables $X_{ij}^{(m)}\equiv \phi_{ij}^{(m)}-\langle\phi_{ij}^{(m)}\rangle$. Here, the average $\langle\cdot\rangle$ may be performed over space (pixels $(i,j)$) or time (frames) according to the required application. If the retrieved phases measure the same surface $h$, the covariance matrix $C$ has a principal component, i.e., one eigenvector $\mathbf{v}_p$ with eigenvalue $\lambda_p$ much larger than the others. In our case, the vector of prefactors $\mathbf{v}$ must then be the principal component of $C$. To evaluate how well the data in $\mathbf{p}$ can be represented by a single quantity $h$ along the direction of $\mathbf{v}$, we observe the normalised eigenvalues $\mathrm{sg}_p\equiv\lambda_p/\sum_k\lambda_k$. We denoted this quantity as the confidence of the PCA, and it takes values from $0$ to $1$, where $\mathrm{sg}_p=1$ implies that all variation in the data is along the principal component $\mathbf{v}_p$. Whereas for $\mathrm{sg}_p=0$, there is no variation in the data.


\section{Results and discussion}
The setup, as shown in Fig.~\ref{fig:holo:schematic}, is proposed to be used as a profile sensor over a circular area on the fluid surface. For this, we employ a diode pumped solid state laser source with wavelength $532$ nm and power of $1$ mW and expand the input beam to our desired width of approximately $20$mm using a beam expander. The fluid sample is placed in a square basin of side $84$~mm with a $1$-inch mirror at its bottom. The expanded probe and reference beams are then set in off-axis alignment as described previously. The multiple reflections from the mirror and the fluid surface give rise to multiple interferograms, in line with~\eqref{eq:holo:image}, and are acquired over time using a grayscale CMOS camera with bit-depth if $12$ bits. Using the procedure presented in section~3\ref{sec:demodulation}, the phases of the available interferograms are then numerically reconstructed. This, in turn, gives information on the fluctuations on the fluid surface, which are mainly due to vibrational noise but also thermal effects, changes in refractive index or other external disturbances. For the purposes of this work, the data was acquired from a setup assembled on a noise-isolating table.
\begin{figure}[!t]
\centering
\includegraphics[width= \linewidth]{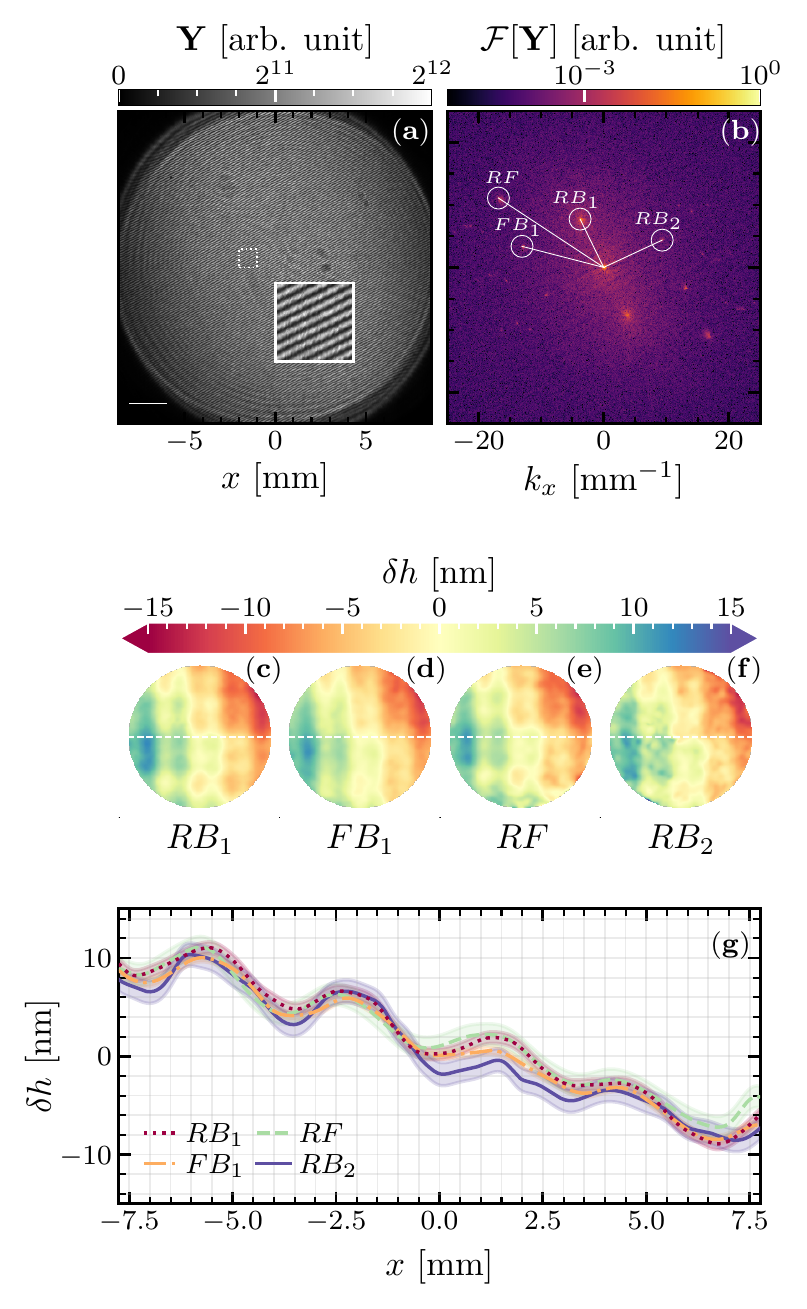}
\caption{\textbf{Reconstruction of water surface profile.} In \textbf{(a)}, a sample image frame acquired from the experimental setup, the solid, white line indicates a length of $2$ mm. The spatial-frequency domain of the image in \textbf{(a)} is shown in \textbf{(b)}. The four main peaks labelled by the procedure described in the text are shown in \textbf{(b)}. Their reconstructed height change obtained by using the inferred prefactors of~\eqref{eq:holo:phasetoheight} are shown in \textbf{(c)}-\textbf{(f)}. The white dashed line in \textbf{(c)}-\textbf{(f)} corresponds to a diameter of $15.5$ mm. In \textbf{(g)}, we display a horizontal slice of $\delta h$ for all labelled peaks. The shaded regions indicate the root-mean-square deviation from the average reconstruction $\delta h = \tfrac{1}{4}\sum_{AB}\delta h_{AB}$ of the four interferograms given by $\sqrt{\langle|\delta h_{AB}-\delta h|^2\rangle_{\mathbf{r}}}$. All curves in \textbf{(g)} qualitatively and quantitatively recover the same interface with various degrees of noise. Albeit small, we see that the spatial shift introduced by the relative tilts between the sample and reference mirrors persists.}
\label{fig:surface}
\end{figure}

\begin{figure}[!t]
\centering
\includegraphics[width=\linewidth]{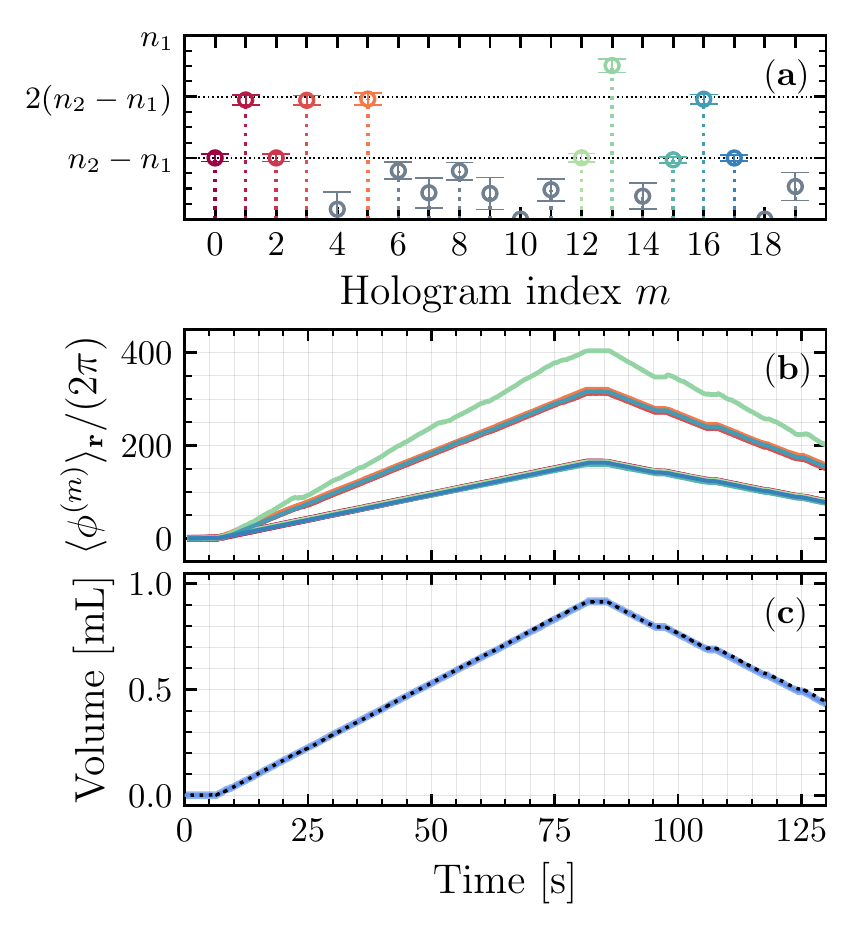}
\caption{\textbf{Fluid depth monitoring.} In \textbf{(a)}, temporal PCA to identify peaks labelled as the hologram indices $m$ and calculate their prefactors. The coloured points indicate the prefactors estimated with a PCA confidence at $97.40\%$. \textbf{(b)} displays the phase changes retrieved from different peaks identified in \textbf{(a)}. In \textbf{(c)}, the blue line indicates the average volume change calculated from different peaks, and the shaded region indicates the RMS deviation of the average from the rest. The dotted line shows the volume measurement done by tracking the movement of the syringe.}
\label{fig:knownvolumedata}
\end{figure}

\subsection{Fluid surface profile \label{exppeakid}}

An example frame $\mathbf{Y}$ is shown in Fig.~\ref{fig:surface}(a) with its spatial Fourier transform $\mathcal{F}[\mathbf{Y}]$ displayed in Fig.~\ref{fig:surface}(b). Unlike the simulated image in Fig.~\ref{fig:holo:syntheticData}, Fig.~\ref{fig:surface}(b) contains many extra dimmer peaks, which are from partial reflections at the faces of the optical components, mainly the beam-splitter. In the experimental Fourier spectrum, it is not straightforward to label the correct peaks associated with different holograms. We developed a graphical user interface with a live view of the spatial frequency domain to aid in the labelling of the various peaks. The reconstructed phase and interferograms can also be viewed in real-time using this application. The most prominent peak in the Fourier space is always the $RB_1$ because of the beams of maximum intensities reflected from the reference and probe mirrors. When the reflected beam from the reference mirror is blocked, the only remaining interference is between the probe mirror and the fluid surface, i.e., holograms of the family $FB_j$s. The brightest one in this group can be labelled as $FB_1$. Further adjustments in the reference mirror alone keep the positions of $FB_j$s fixed, and with the adjustments in the probe mirror alone, the position of $RF$ remains unaffected. This method allows for the preliminary identification and labelling of the prominent peaks from the spatial-frequency domain of the experimental data. 

Upon identifying the peaks and hence their phase-prefactors according to~\eqref{eq:holo:phasetoheight}, the height fluctuations $\delta h(\mathbf{r},t)$ on the fluid surface can be determined using the reconstructed phases. In Fig.~\ref{fig:surface}(c-f), we display the height change $\delta h$ reconstructed from multiple peaks between two consecutive frames $12.5$~ms apart, i.e., $\delta h \equiv h(\mathbf{r},t+\delta t)-h(\mathbf{r},t)$. The panels (c-f) in Fig.~\ref{fig:surface} are ordered by decreasing intensity of their corresponding peak, i.e., $RB_1$ is the brightest and $RB_2$, the faintest. It is clear that all four independent reconstructions retrieve the same qualitative profile of the surface. However, as evident in Fig.~\ref{fig:surface}(g), the quality of the reconstruction varies amongst the peaks. For instance, since $RB_2$ has low intensity, its signal is more likely to be overwhelmed by the background noise level, resulting in a noisier reconstruction, as can be seen in Fig.~\ref{fig:surface}(f) and by the dashed, blue curve in  Fig.~\ref{fig:surface}(g).

\subsection{Depth change in fluids}

Digital holography is a promising technique to monitor the dynamic changes in the probe arm. In the previous section, we showed that it is possible to resolve relative temporal changes down to nanometric scales. However, we offer an alternative analysis for identifying the peaks through data processing of the acquired images.  Here, the principles of our modelling are used to evaluate the time evolution of the fluid depth with respect to a reference time. For the experiments that follow, we coarsely aligned the reference and sample mirrors to demonstrate the potential application of our methods in scenarios where careful alignment cannot be guaranteed, such as for usage in noisy environments. Consequently, the spatial frequency domain of the acquired data may display common signal processing faults, such as spatial Fourier harmonics and folding, repeated peaks and aliasing~\cite{sundararajan2017digital}. Due to these digital artefacts, it is also expected that, in this case, not all peaks will be present or be identified in the process.

For the following analysis, we note that the overall change in time of the sample's depth is directly proportional to the change in the spatial average of the phases, i.e., $\langle \delta h(\mathbf{r},t)\rangle_{\mathbf{r}}\propto\langle \delta\phi^{(m)}(\mathbf{r},t)\rangle_{\mathbf{r}}$, for a peak $m$. The latter can be inferred from the complex value of the peak alone, as it carries the spatial average in its phase $\phi^{(m)}$~\cite{sundararajan2017digital}. Thus, by employing the phase recovery procedure of~\eqref{eq:syntheticReconstruction} to all peaks in the Fourier spectrum, we can obtain the change in the spatially averaged phases. We can then apply the PCA procedure presented in section $3$\ref{peakid} to identify the phase prefactors that give the appropriate relative fluid depth.

\subsubsection{Controlled change in volume}
In order to confirm that our setup can be used to reliably monitor the change in depth of the fluid sample, we devised an experiment where the volume of fluid in the basin was varied in a controlled way and monitored independently. A known volume change $dV$ is injected into the sample basin, and the corresponding phase change $d\phi^{(AB)}$ is given by,
\begin{equation}
    d\phi^{(AB)} =\frac{dV}{2k_0 \alpha_{AB} A}
\end{equation}
where $A$ is the area of the cross-section of the basin, which for this experiment was $70.6(2)~\mathrm{cm}^2$. The height fluctuations $h(\mathbf{r},t)$ are calculated from the reconstructed phases $\phi$ and are then compared with the predicted change in volume. The fluid level in the basin was changed steadily by employing a syringe system attached to a remote-controlled stepper motor. Water is injected into the basin at a rate of $12.3$ $\mu$L/s. Fig.\ref{fig:knownvolumedata} (c) and Fig.\ref{fig:knownvolumedata} (b) show the rate of change in the volume of the fluid over $130$ seconds and the corresponding phase change for different holograms. 

From the temporal PCA, it was found that the phases from different holograms were consistent towards a single principal component with confidence of $97.40\%$, which shows the reliability of the method of reconstruction. The phase prefactors $\alpha_{AB}$ obtained from the analysis are given in Fig.\ref{fig:knownvolumedata} (a). Similar to what was discussed above, the phase prefactor obtained for the holograms 0, 2, 12, 15 and 17 is $n_2-n_1$, which indicates that all of them correspond to the same height variation as $RB_1$ and repetitions of its peak. The holograms 1, 3, 5 and 16, with prefactors $2(n_2-n_1)$, are consistent with the first harmonics of the $RB_1$ peak and its repetitions, and they are artefacts of digital processing. We could not identify confidently peak 13 within our model, similar to those in Fig~\ref{fig:evaporationwaterdata} (c). The holograms shown in grey in Fig.\ref{fig:knownvolumedata} (a) with a prefactor less than $n_2-n_1$ are not used further as they are not well correlated to the height field $h(\mathbf{r},t)$. 

\subsubsection{Evaporation Rates} 

\begin{figure}[!t]
\centering
\includegraphics[width= \linewidth]{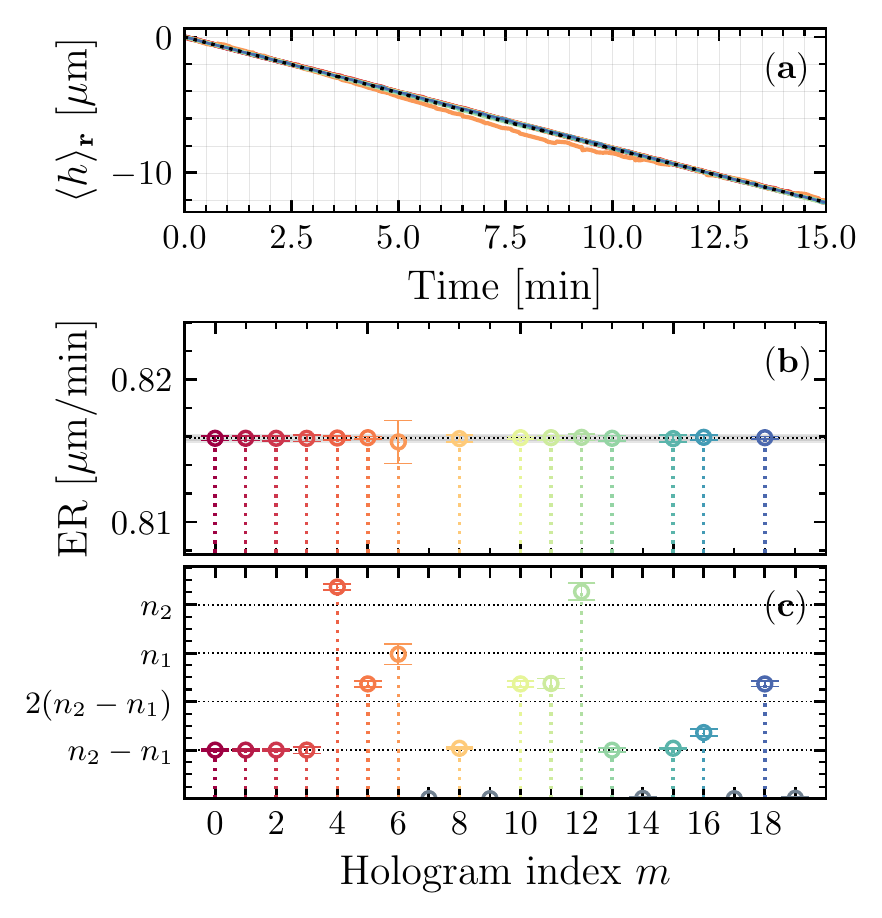}
\caption{\textbf{Evaporation rate of Water.} In \textbf{(a)}, the instantaneous height fluctuations retrieved from the reconstructed phases of multiple peaks and the dotted line gives the average of them. In \textbf{(b)}, the linear fit of the average depth change in \textbf{(a)} and thus, the evaporation rate is depicted as the dotted line, the coloured points give the evaporation rates estimated from linear fits of each of the relevant peaks. The evaporation rate extracted from the slope of the PCA averaged depth is $0.8159(2)~\mu\mathrm{ m/min}$. From Hisatake et al. in~\cite{hisatake1995experimental} and~\cite{hisatake1993evaporation}, for low air-current speeds and similar ambient temperatures and water surface area, we should expect evaporation rates in the range of $0.6-1.2~\mu\mathrm{m/min}$. The temperature of the experiment was at $20.8(2)^{\circ}\mathrm{C}$. The peak identification and prefactor estimates of different peaks using temporal PCA with confidence at $99.94\%$ in shown in \textbf{(c)}. For this data set, data were acquired at a rate of $2$ frames per second.
    }
\label{fig:evaporationwaterdata}
\end{figure}
\begin{figure}[!t]
\centering
\includegraphics[width= \linewidth]{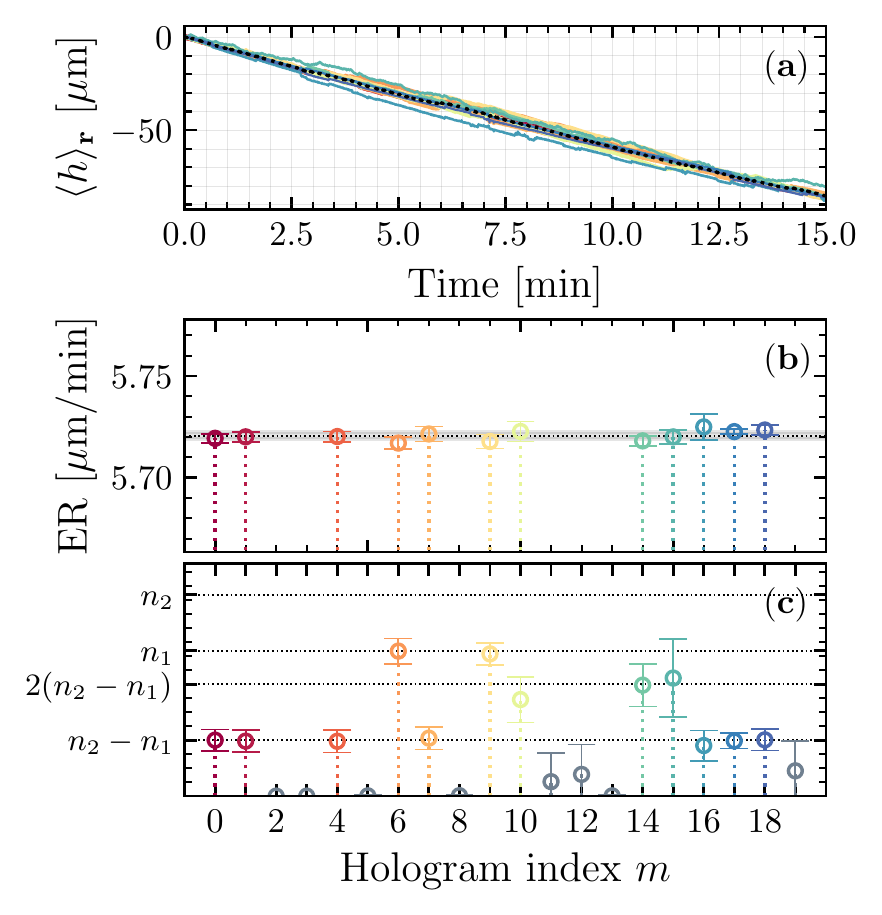}
\caption{\textbf{Evaporation rate of Isopropanol.} Panels \textbf{(a)} and \textbf{(b)} display the same content as in Fig~\ref{fig:evaporationwaterdata}, but for isopropanel as the fluid sample. The evaporation rate extracted from the slope of the PCA averaged depth is $5.720(1)~\mu\mathrm{ m/min}$. From Mackay and van Wesenbeeck~\cite{mackay2014correlation} and references therein, we can also estimate rates in the range of $5-6~\mu\mathrm{m/min}$. The temperature in the surroundings of the experiment was 
at $23.2(2)^{\circ}\mathrm{C}$. In \textbf{(c)}, we show the peak identification and prefactor estimates of different peaks using temporal PCA with confidence at $99.25\%$. The hologram index $m$ is the order of peaks identified in the Fourier space. For this dataset, images were acquired at a rate of $10$ frames per second.    }
\label{fig:evaporationIsopropanoldata}
\end{figure}

In this study, the sample fluid in the fluid cell was allowed to remain idle for $15$ minutes on a noise-isolating table in a controlled environment, with negligible air currents and variations in temperature. Fig.~\ref{fig:evaporationwaterdata} and Fig.~\ref{fig:evaporationIsopropanoldata} show the results of the temporal PCA done on the data acquired using water ($n_2 = 1.333$) and isopropanol ($n_2 = 1.377$), respectively, as the samples. Panels (a) in both figures depict the various reconstructed depth changes over time for the available peaks with respect to the initial position, set to zero for convenience. We used the root-mean-square deviation of each curve with respect to the average (dotted, black line) as a measure of uncertainty for the prefactors obtained through the PCA. With water, the PCA confidence was at $\mathrm{sg}_p=99.94\%$, which confirms that the different holograms correspond to the same depth change trend with a high level of accuracy.  

Fig.\ref{fig:evaporationwaterdata} (c) shows the phase-prefactors $\alpha_{AB}$ obtained from the analysis using water as the sample. The holograms are numbered in descending order of their intensities. Thus, hologram $0$, which is the most prominent (brightest) one, corresponds to $RB_1$ and hence its prefactor can be identified as $n_2-n_1$. The same phase prefactor was obtained for the holograms 1, 2, 3, 8, 13, and 15, which indicates that all of them correspond to the same height variation as $RB_1$ and are likely to have come from repetitions of this peak. Holograms 4 and 12 have the highest phase prefactor among all the peaks. With the pre-factor close to $n_2$, allowing to identify them as the $FB_1$ hologram and a repeated peak. The hologram 6 with a prefactor $n_1$ represents the $RF$. The other holograms shown in the figure are either weakly correlated to the height changes $\delta h(\mathbf{r},t)$ (grey points) and can be discarded, or do not correspond to peaks predicted by our modelling, such as interference with other optical components (e.g., the beam splitter), but still result in a hologram of the fluid surface. Similar conclusions can be derived from the data using isopropanol as shown in Fig.~\ref{fig:evaporationIsopropanoldata}, with a PCA confidence of $99.25\%$. In both experiments, the evaporation rates (ER) of the fluids were estimated from the linear fit of the various depth change curves in Panels (a), and the values were found to agree with those from relevant literature (see captions of Fig.~\ref{fig:evaporationwaterdata} and Fig.~\ref{fig:evaporationIsopropanoldata}). 

\section{Conclusion}

In this work, we presented an off-axis DH setup that achieves multiplexing by considering the multiple reflections from the transparent fluid sample and other optical components. The multitude of holograms embedded in the interfered intensity image can then be separated into individual contributions that appear as brightness peaks in the spatial frequency domain (SFD) of the image. Phase modulations around each of these peaks are directly proportional to the height of the sample. We showed that a simple optical configuration based on a Michelson interferometer, when combined with a refined analysis of the acquired data, can retrieve a reliable reconstruction of spatial and time-dependent variations of the surface profile of the sample. 

By carefully aligning the relative tilt between the reference and basin mirrors, our current setup can resolve nanometric changes in a water-air interface (see Fig.~\ref{fig:surface}). The phases of different holograms resulted in the same reconstructed profile with varying noise. We further showed that the position of the first four brightest peaks in SFD is in line with the prediction of our model in~\eqref{eq:holo:interferogramsPhases}. Thus, by identifying two of the peaks, the position of the remaining ones can be inferred from the appropriate combination of the position vectors in SFD using~\eqref{eq:holo:interferogramsPhases}.   This method could be used to investigate the dynamics of fluid flows and the dispersion of interfacial waves with improved precision than conventional methods in Fluid Profilometry~\cite{Wildeman2018}.

We then offered an alternative statistical treatment of the data to target scenarios where the alignment may not be accurately done. By employing a principal component analysis on the spatial averages of various phases, we showed that it is possible to precisely monitor changes in the depth of the sample when the volume of fluid is slowly varied in a known fashion. The same principles used in this benchmark were extended to investigate the evaporation rate of liquids. Our results are in accordance with the literature on the subject and confirm a possible application for the method. By trading the careful alignment for statistical treatment, it is not always possible to identify all the expected holograms, and the $RB_1$ family seems to be dominant in all cases presented here.

Our method heads towards the development of profilometry sensors with uncomplicated designs relying strongly on digital processing and statistical tools to deliver better performance. The quality of the reconstructed profiles can be improved and adapted to different samples by modifying specific components, such as the camera, the laser source, or the optical parts, granting our approach a modular feature.  With fast-evolving machine learning algorithms and artificial intelligence favouring automation, further investigation includes combining automatic alignment of the setup with appropriate statistical treatment of the data to optimise the result of the reconstruction in a wide range of applications.

\paragraph*{Acknowledgements.}
We thank the members of the Gravity Laboratory and the Quantum Simulators for Fundamental Physics extended community for fruitful discussions and helpful suggestions. We are grateful to the Technology Transfer Office team and the technical support team of the Physics \& Astronomy department at the University of Nottingham for their continued support in the development of this work.
SW acknowledges support provided by the Leverhulme Research Leadership Award (RL-2019- 020),
the Royal Society University Research Fellowship
(UF120112, RF\textbackslash ERE\textbackslash 210198) and the Royal Society Enhancements Awards and Grants
(RGF\textbackslash EA\textbackslash 180286, RGF\textbackslash EA\textbackslash 181015), and partial
support by the Science and Technology Facilities Council (Theory Consolidated Grant ST/P000703/1), the Science and Technology Facilities Council on Quantum Simulators for Fundamental Physics (ST/T006900/1) as part
of the UKRI Quantum Technologies for Fundamental
Physics programme.

\paragraph*{Disclosures.} AG, VSB and SW: University of Nottingham, UK Patent Application Number \href{https://www.ipo.gov.uk/p-ipsum/Case/ApplicationNumber/GB2214343.2}{GB2214343.2} (P).

\paragraph*{Data availability.} Data underlying the results presented in this paper are not publicly available at this time but may be obtained from the authors upon reasonable request.

\bibliographystyle{ieeetr}
\bibliography{bibfile}



\section*{Supplementary material}

# Multiplexed digital holography for fluid surface profilometry: Supplemental document

This supplementary material constitutes the basic theoretical framework and the analytical techniques used in the simulation and data analysis.

## I. THEORETICAL CALCULATION OF THE PHASES AND THE AMPLITUDES

Consider a collection of (complex) propagating electromagnetic waves that are traversing a medium, which is considered to be homogeneous on length scales comparable to the wavelength $\lambda$. Within the geometrical optics limit, the electric field strengths can be written as,

$$\mathbf{E}_n = E_0 \mathbf{A}_n \exp\left(i\Phi_n - i\omega t\right), \quad \text{for} \quad \Phi \equiv \int \mathbf{k} \cdot d\mathbf{r} . \tag{1}$$

Here $\omega$ is the angular frequency of the ray, and $\mathbf{A}_n$ is a complex, dimensionless vector directed along the axis of polarization. Its magnitude $|\mathbf{A}_n|$ gives the relative amplitude with respect to the electric field amplitude $E_0$. For a superposition $\{\mathbf{E}_n\}$ of rays of the form (1) at the same frequency $\omega$, the (time-averaged) intensity is given by,

$$I(t, \mathbf{r}) \equiv I_0 + I_0 \sum_{n \neq m} A_{nm}(t, \mathbf{r}) e^{i\Phi_{nm}(t, \mathbf{r})}, \tag{2}$$

for $\mathbf{A}_n^\dagger \mathbf{A}_m \equiv A_{nm}$ and $\Phi_{nm} \equiv \Phi_n - \Phi_m$. Here, and in what follows, the conjugate transpose is denoted by a superscript $\dagger$. The average total intensity is $I_0 = \frac{1}{2}\varepsilon c |E_0|^2$, where $\varepsilon$ and $c$ are the medium's electric permittivity and speed of light, respectively. Hence, $|E_0|$ is defined through the relation $\sum_n |\mathbf{E}_n|^2 \equiv |E_0|^2$ or, equivalently, $\sum_n |\mathbf{A}_n|^2 = 1$.

From equation (2), it follows that for any two distinct rays $n$ and $k$ with unequally varying phases ($\Phi_{nk} \neq 0$ and $|\nabla \Phi_{nk}| \neq 0$), the intensity $I$ must fluctuate locally in space. Due to the short wavelengths of light, the resulting intensity distributions, commonly referred to as interferograms or interference patterns, are highly sensitive to relative changes along or of the paths travelled by the interfering beams.

Consider that the beam splitter divides the laser beam into reference ($R$) and probe ($P$) beams. Let $z$ be the direction of the incident beam. The probe beam is now illuminating the fluid surface, $z = h(t, \mathbf{r})$, where h = constant, when the fluid is at rest. Here, t is time and $\mathbf{r} = x\mathbf{e_x} + y\mathbf{e_y}$ are the coordinates for the horizontal plane. A plane mirror, $z = -\mathbf{m}_b \cdot \mathbf{r}$ is submerged under the interface. Here, $\mathbf{m}_b$ is the angle of rotation of the mirror from the optical axis, assuming that the tilt is small, and $\mathbf{m}_b = 0$ corresponds to the submerged mirror being parallel with the undisturbed fluid interface $z = $ constant. The water in the basin acts as a cavity to the incoming ray P, and gives rise to multiple reflections and transmissions. A ray that passes through an infinitesimal distance $dz$ of an isotropic medium with refractive index $n$ attains a phase $d\Phi = k_0 n dz$, where $k_0 2\pi/\lambda_0$ is the wave number of the ray in vacuum. The additional reference ray, labelled $R$, is reflected from the mirror $M_r$ but with a different tilt, $\mathbf{m}_b \to \mathbf{m}_r$, and a different arm length, $H \to H_r$.

We shall now consider the phases accumulated by different types of rays as they traverse a fluid interface $z = h(t, \mathbf{r})$. Let $F$ be the beam reflected from the surface of water and $B_j$'s, the ones reflected from the submerged mirror. The amplitudes of these beams are, then, given by,

$$A_R = \frac{1}{4}, \tag{3a}$$

$$A_F = \frac{1}{4}\mathcal{R}, \tag{3b}$$

$$A_{B_j} = \frac{1}{4}\mathcal{R}^{j-1}(1 - \mathcal{R}^2), \tag{3c}$$

Here, $\mathcal{R} \equiv \frac{n_2 - n_1}{n_1 + n_2}$ is the reflection coefficient of water-air interface and $n_1$, $n_2$, the refractive indices of air and water, respectively. The factor $\frac{1}{4}$ in the equations above is due to the beams propagating through the beam splitter multiple times. It comes from the factor of $R_B T_B$ that should be applied to the amplitudes if the intensity $I_0$ is to refer to the intensity prior to passing a beam splitter with transmission coefficient $T_B$ and reflection coefficient $R_B$. The corresponding phases are given by,

$$\Phi_R = 2kn_1(z_0 - \mathbf{m}_r \cdot r), \tag{4a}$$

$$\Phi_F = 2kn_1(z_0 - h), \tag{4b}$$

$$\Phi_{B_j} = 2kn_1(z_0 - h) + 2kjn_2(h - \mathbf{m}_b \cdot r), \tag{4c}$$

The interferograms used to reconstruct the surface fluctuations in the media are from the interferences of reference beam R with the surface reflection F and multiply reflected beams $B_j$'s, and those between F and $B_j$'s, leading to various peaks in the Fourier spectrum. The interferograms can now be expressed as:

$$\textbf{Interferogram}(\textbf{RF}) \rightarrow A_{RF}\, cos(\Phi_{RF}), \tag{5a}$$

$$\textbf{Interferogram}(\textbf{RB}_\textbf{j}) \rightarrow A_{RB_j}\, cos(\Phi_{RB}), \tag{5b}$$

$$\textbf{Interferogram}(\textbf{FB}_\textbf{j}) \rightarrow A_{FB_j}\, cos(\Phi_{FB}), \tag{5c}$$

$$\textbf{Interferogram}(\textbf{B}_\textbf{l}\textbf{B}_\textbf{j}) \rightarrow A_{B_l B_j}\, cos(\Phi_{B_l B_j}), \tag{5d}$$

where $A_{RF}, A_{RB}, A_{FB_j}$ and $A_{B_l B_j}$ are the effective amplitudes and $\Phi_{RF}$, $\Phi_{RB_j}$, $\Phi_{FB_j}$ and $\Phi_{B_l B_j}$ are the effective phase differences from the interferences.

The phase differences are given by

$$\Phi_{RF} = 2k_0 n_1 \left(h + \textbf{m}_r \cdot \textbf{r} + \Delta H\right), \tag{6a}$$

$$\Phi_{RB_j} = 2k_0[(n_1 - jn_2)h + (n_1\textbf{m}_r - jn_2\textbf{m}_b) \cdot \textbf{r} + n_1\Delta H], \tag{6b}$$

$$\Phi_{FB_j} = 2k_0 n_2[jh + j\textbf{m}_b \cdot \textbf{r}], \tag{6c}$$

$$\Phi_{B_\ell B_j} = 2k_0 n_2[(\ell - j)h + (\ell - j)\textbf{m}_b \cdot \textbf{r}], \tag{6d}$$

with $\Delta H \equiv H_R - H$ the difference in arm length of the reference beam $H_R$ and the probe beam $H$. While the corresponding amplitudes take the form

$$A_{RF} = \frac{1}{8}\mathcal{R}, \tag{7a}$$

$$A_{RB_j} = \frac{1}{8}(1 - \mathcal{R}^2)\mathcal{R}^{j-1}, \tag{7b}$$

$$A_{FB_j} = \frac{1}{8}(1 - \mathcal{R}^2)\mathcal{R}^j, \tag{7c}$$

$$A_{B_\ell B_j} = \frac{1}{8}(1 - \mathcal{R}^2)^2 \mathcal{R}^{\ell+j-2}, \tag{7d}$$

Because $|\mathcal{R}|$ is typically small ($|\mathcal{R}| \simeq 14.16\%$ for air-water interfaces), the $RB_1$ interferogram has the largest amplitude. The sensitivity of the phase to variations in the height, $\partial_h \Phi_{RB_1} = 2k_0(n_1 - mn_2)$, however, is the smallest. The most phase-sensitive of the $j = 1$ interferograms is $FB_1$, with $\partial_h \Phi_{FB_1} = 2k_0 n_2$. Crucially, the intensity contributed by the three interferograms $RF$, $RB_1$ and $FB_1$ have distinct external influences. For example, inhomogeneities in the bulk of the surrounding gas, from changes in the composition of the gas, the temperature, or from sound waves, are picked up by $RF$ and $RB_1$, but not by $FB_1$, which only probes the fluid bulk. For transparent fluids, $FB_1$ is expected to have the largest signal-to-noise ratio, provided the reflection coefficient $|\mathcal{R}|$ is large enough so its intensity is detectable.

We can then consider the spatial Fourier spectrum of the combined intensity (2) for phases given by (6). By virtue of the distinct tilts $\textbf{m}_r$ and $\textbf{m}_b$ of the reference and bottom mirrors respectively, the interferograms separate into distinct spatial carrier frequencies (see Fig. 1). Their separation in spatial-frequency domain is what guarantees that our methods presented in the main text can be applied.

## II. SIMULATION WITH NUMERICAL PHASE TRACING

The numerical phase tracing is based on the iterative propagation of a mesh of rays through a synthetically generated model of the setup. More precisely, we represent the two mirrors as tilted planes $z = -\textbf{m}_b \cdot \textbf{r}$ and $z = -\textbf{m}_r \cdot \textbf{r}$, and then the interface $z = h(\textbf{r})$ takes the form:

$$h(\textbf{r}) = h_0 + A[\cos(\textbf{k}_1.\textbf{r}) + \cos(\textbf{k}_2.\textbf{r})], \tag{8}$$

Now, consider a cartesian mesh of starting positions for the initial ray, which is a set of two-dimensional points $\textbf{r}_{ij} \in \mathbb{R}^2$ in the (x,y)-plane, at a fixed altitude $z = z_0$. The origin of the rays is then given by $\textbf{x}_{ij} = z_0\textbf{d}_{ij} + \textbf{r}_{ij}$, where $\textbf{d}_{ij} = -\textbf{e}_z$ is the initial direction of the rays. The initial phase and relative amplitude associated with the rays are taken as $\phi_{ij} = 0$ and 1, respectively.

The refracted mesh of rays initiates new rays at each reflection of the free surface from below. At every interface encountered, the amplitudes are modified, and at every propagation of length $L$ between interfaces, the phase attains a contribution $kL = k_0 nL$, where $k_0$ is the wave number of the ray in vacuum and $n$ is the refractive index of the traversed medium.

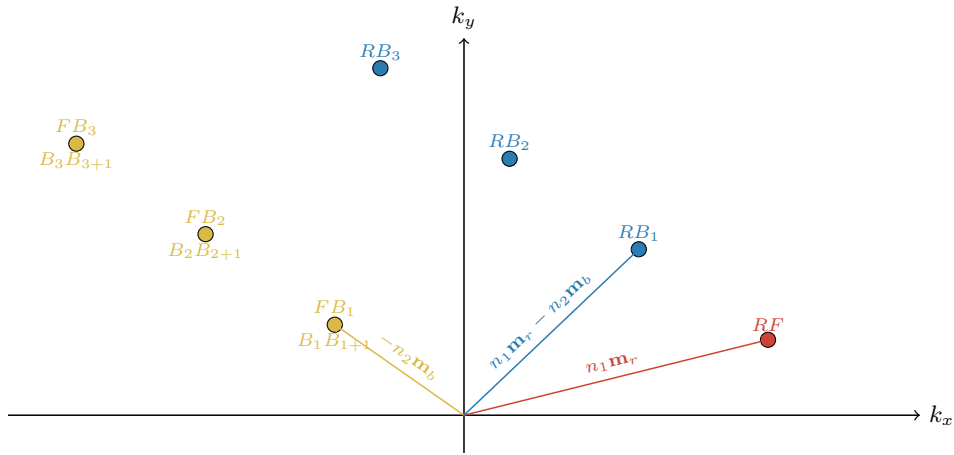

FIG. 1. **(Separation of Carrier Peaks)** The separation of spatial carrier frequencies $\mathbf{k} = (k_x, k_y)$ that results from small tilts $\mathbf{m}_b$ and $\mathbf{m}_r$ of the submerged mirror and reference mirror respectively. Each node represents a hologram specified by the label. The holograms are categorised into three groups (yellow, blue and red). Yellow holograms are independent of the configuration of the reference mirror $\mathbf{m}_r$, the red hologram is independent of the submerged mirror $\mathbf{m}_b$ and the blue nodes respond to adjustments in both mirrors.

**Step 1:** For each mathematical surface $F_k(x)$, numerically solve the equation, $F_k(x_{ij} + s_k d_{ij}) = 0$, for $s_k$, and determine the $k = k_0$ with smallest non-negative value s for $s_k$, and label $F_{k_0}$ as the current surface. If there are no solutions $s_k$, choose $s = (z_0 - x_{ij} \cdot \mathbf{e}_z)/(\mathbf{d}_{ij} \cdot \mathbf{e}_z)$ to propagate the ray back to $z_0$, and move to step (4) after this step. Update the position $x_{ij} \longrightarrow x_{ij} + s d_{ij}$ and the phase $\phi_{ij} \longrightarrow \phi_{ij} + k_0 n_{ij} \mathbf{s} |\mathbf{d}_{ij}|$.

**Step 2:** Compute the normal vector $n \equiv \nabla F_{k_0}/|\nabla F_{k_0}|$ of the current surface and find the reflected $\mathbf{d}_{ij}^r$ and refracted $\mathbf{d}_{ij}^t$ directions about the surface normal using Snell's law, i.e.

$$\mathbf{d}_{ij}^r = 2c\mathbf{n} \text{ and } \mathbf{d}_{ij}^t = n_r \mathbf{d_{ij}} - \left(n_r c + \sqrt{1 - n_r^2(1 - c^2)}\right)\mathbf{n}, \qquad (9)$$

where $c = \mathbf{n} \cdot \mathbf{d}_{ij}$ and $n_r = n_a/n_b$ is the current refractive index $n_a$ divided by the refractive index $n_b$ behind the surface. Duplicate the current ray, assign the (reflected) direction $\mathbf{d}_{ij}^r$ to the duplicate, and reduce the amplitude by the Fresnel reflection coefficient, i.e. $A_{ij} \longrightarrow \mathcal{R}_{ab}A_{ij}$. Pass the duplicate to step (3). Update the refractive index $n_a \longrightarrow n_b$, direction $\mathbf{d}_{ij} \longrightarrow \mathbf{d}_{ij}^t$ and amplitude $A_{ij} \longrightarrow \tau_{ab}A_{ij}$ of the current ray and pass it to step (3). Here, $\tau_{ab}$ is the Fresnel transmission coefficient.

**Step 3:** If the amplitude is sufficiently large, i.e. $|A_{ij}| > \epsilon$ for some $\epsilon$ (the chosen value here is $\epsilon = 5$ x $10^{-4}$), send the ray back to step (1). If not, discard the ray.

**Step 4:** Wait for all rays to reach this step. When they do, move to step (5).

**Step 5:** Use the (x, y)-coordinates of the current position $x_{ij}$ to interpolate the amplitudes $A_{ij}$ and phases $\phi_{ij}$ to the original transverse coordinate $\mathbf{r}_{ij}$. Then use (5a) for rays $A_{ij}e^{i\phi_{ij}}$ to compute the intensity $I_{ij}$. The value of the resulting image $Y_{ij}$ at pixel (i, j) is chosen by sampling from a Poisson distribution with mean $\gamma I_{ij}$. Here $\gamma$ is the average number of photons arriving in the camera during the exposure time, we chose it to be 800.

## III. DEMODULATION OF NOISY CARRIERS

Consider an image $Y_{ij}$ created from intensities of the form (2), $Y_{ij} = \Gamma[I(\mathbf{r}_{ij})] + \delta Y_{ij}$ where $\delta Y_{ij}$ is a stochastic variable that describes the noise in the image, $\mathbf{r}_{ij}$ is the spatial (pixel) coordinate, and $\Gamma$ is a function that transforms physical intensities to pixel values. We shall assume all phase differences $\Phi_{mk} = \Phi_m - \Phi_k$ to have small variations $\phi$ around a stationary planar phase $\mathbf{k} \cdot \mathbf{r}$. That is, $\Delta\Phi(t,\mathbf{r}) = \phi(t,\mathbf{r}) + \mathbf{k} \cdot \mathbf{r}$, so that

$$Y_{ij} = Y_0 + \sum_n B_n \exp\left[i\mathbf{k}_m \cdot \mathbf{r}_{ij} + \phi_{ij}^{(m)}\right] + \delta Y_{ij} \equiv \text{Re}[\tilde{Y}_{ij}] + \delta Y_{ij}, \qquad (10)$$

for $B_n \equiv Y_0 A_m$ and $Y_0 \equiv \Gamma I_0$, the pixel value corresponding to the spatially averaged intensity $I_0$. The equation above shows that the Fourier spectrum of the image $Y_{ij}$ displays intensity peaks around the $\mathbf{k}_m$ vectors. Their positions in the frequency domain are defined in accordance with the wavevectors in the planar phases. If the peaks are sufficiently separated, one can identify them as in figure 1. Since $Y_{ij}$ is real, each non-zero carrier $\mathbf{k}_m$ must have a so-called *twin* at $-\mathbf{k}_m$.

Now let $F_m$ be a Fourier filter around a single carrier peak $\mathbf{k}_m$, i.e. $F_m \equiv \mathcal{F}^{-1} G_m(\mathbf{k}) \mathcal{F}$, where $\mathcal{F}$ is a two-dimensional spatial Fourier transform with inverse $\mathcal{F}^{-1}$, and $G_m$ is a window function that is only non-zero in the vicinity of the $m$'th carrier peak $\mathbf{k}_m$. The role of the filter $F_m$ is to isolate each modulated plane wave to get,

$$F_m[\tilde{Y}]_{ij} = B_m \, \exp\left[i\mathbf{k}_m \cdot \mathbf{r}_{ij} + \phi_{ij}^{(m)}\right] + F_m[\delta Y]_{ij}, \tag{11}$$

The simplest choice for $G_m(\mathbf{k})$ is a two-dimensional tophat-filter centred at $\mathbf{k}_m$, i.e. $G_m(\mathbf{k}) = 1$ if $|\mathbf{k} - \mathbf{k}_m| \leq k_c$ with $G_m(\mathbf{k}) = 0$ otherwise, for some appropriate choice of filter size $k_c$. However, the reconstruction can be significantly improved by using a Raised-Cosine filter. Using the Fourier filter $F_n$, we have extracted a complex signal whose only deviation from a complex pane wave, other than the noise, is due to the spatial variations of $\phi_{ij}^m$. Assuming the noise to be much smaller than the signal in each peak, i.e. $F_m[\delta Y]_{ij} \ll \tilde{Y}_{ij}$, the complex phase of the filtered image $F_m[Y]_{ij}$ takes the form,

$$\mathrm{Im}\left\{\log\left[F_m[Y]_{ij}\right]\right\} = \mathbf{k}_m \cdot \mathbf{x}_{ij} + \phi_{ij}^{(m)} - \mathrm{Re}\left[\delta\phi_{ij}^{(m)}\right] \ \text{for} \ \delta\phi_{ij}^{(m)} \equiv i\frac{F_n[\delta Y]_{ij}}{\tilde{Y}_{ij}^{(m)}}, \tag{12}$$

for some integer field $\ell_{ij}^{(m)} \in \mathbb{Z}$ enumerating the number of phase-wrappings at the pixel location $(i, j)$. Note that most of the quantities on the right side of (13) are accessible from the image $Y_{ij}$. The carrier peak $\mathbf{k}_m$ is found from the peaks of $|\mathcal{F}Y|$ and is used to construct the filter $F_m$, and the phase wrapping $\ell_{ij}^{(m)}$ can be reduced to a single integer $\ell_m \in \mathbb{Z}$ using e.g. the algorithm of [1]. This leaves only one scalar quantity, the global phase wrapping $\ell_n$, in addition to the noise $\delta\phi_{ij}^{(m)}$, that is yet to be determined.

The phases can now be computed in different ways - synthetic, relative or absolute reconstruction as per the requirements. Assuming that there is no variation in the carrier $k_m$, the phase variations $\Delta\phi_{ij}^{(m)}$ are then given by,

$$\Delta\phi_{ij}^{(m)} = \mathrm{Im}\log(F_m[Y]_{ij}) - \mathbf{k}_m \cdot \mathbf{r}_{ij}. \tag{13}$$

This method is called synthetic reconstruction in which we used a synthetic wavefront as a reference for the reconstruction.

Relative reconstruction involves estimating the phases by considering pairs of consecutive interferograms $[Y_m]_{ij}$ and $[Y_{m+1}]_{ij}$ such that

$$\Delta\phi_{ij}^{(m)} = \mathrm{Im}\left\{\log\left[(F_m[Y_m]_{ij})(F_m[Y_{m+1}]_{ij})^*\right]\right\}, \tag{14}$$

where $*$ denotes the complex conjugate. In the third method, absolute reconstruction, the phases are calculated by taking the first interferogram $[Y_0]_{ij}$ as the reference, as follows,

$$\Delta\phi_{ij}^{(m)} = \mathrm{Im}\left\{\log\left[(F_m[Y_m]_{ij})(F_m[Y_0]_{ij})^*\right]\right\}. \tag{15}$$

In figure 2, we display an example of phase distortion from a jet of compressed air, which causes variations in temperature, pressure and composition of the bulk.

## IV. PEAK IDENTIFICATION AND PREFACTOR ESTIMATES USING PRINCIPAL COMPONENT ANALYSIS

As already discussed, one could use different approaches to identify and label the peaks in the Fourier space. In this section, we discuss the peak identification using the Principal Component Analysis (PCA) [2, 3]. PCA is a statistical tool that can be used when multiple features of a data set correspond to the same observable in the experiment. Here, the relative phases reconstructed from different peaks correspond to the same height fluctuations $\delta h(r, t)$ on the fluid surface.

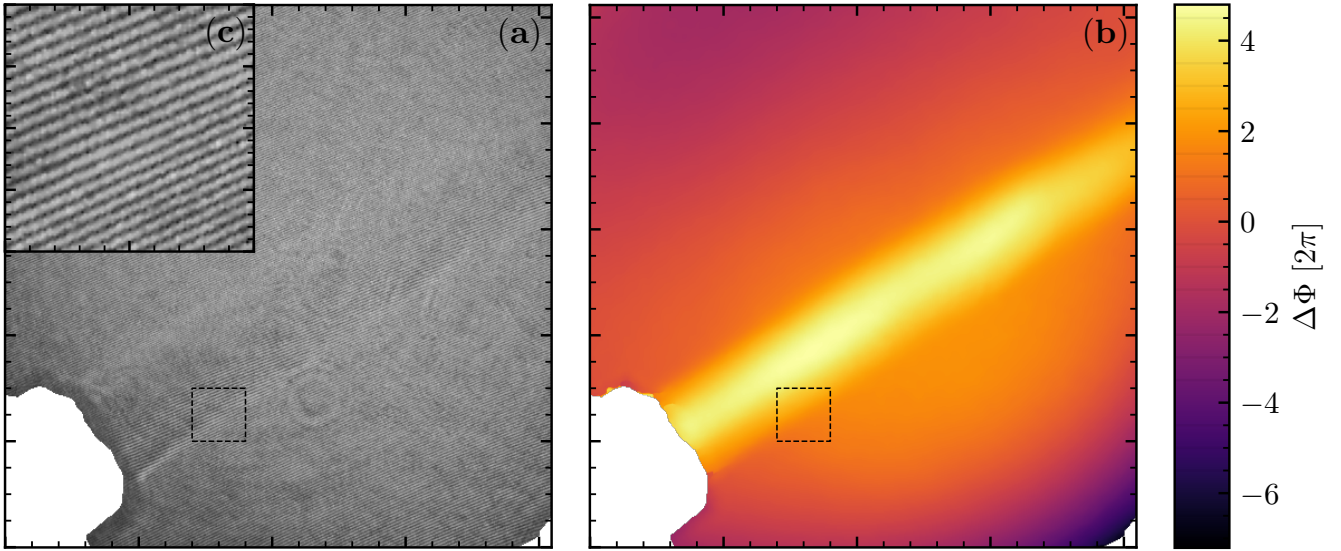

FIG. 2. A jet of compressed air leaving the tip (white region) of an air duster. The jet modifies the local composition $\{\gamma_i\}$ of the air, and also its temperature and pressure, which results in local deformations of the hologram panels (a) and (c), which, when reconstructing the $RB_1$ phase (panel (b)), leaves a clear profile of the density of the jet.

Assume that we have a collection of $M$ reconstructed phases $\{\phi_{ij}^{(m)}\}_{m=1}^M$ that can be seen as an $M$-dimensional vector field,

$$\mathbf{p} = \sum_m \phi_{ij}^{(m)}(t, \mathbf{r})\,\mathbf{e}_m. \tag{16}$$

A spatial or temporal fluctuation of the height $h \to h + \delta h$ results in a fluctuation $p \to p + 2k_0\mathbf{v}\delta h$ for a constant vector,

$$\mathbf{v} = \sum_m \alpha_m \mathbf{e}_m \tag{17}$$

which depends only on the refractive indices. The vector $\mathbf{p}$ always responds to variations in $h$ along the same axis of $\mathbf{v}$, and this axis determines all the phase ratios. Consider the spectrum $C\mathbf{v}_k = \lambda_k \mathbf{v}_k$ of the covariance matrix,

$$C_a b = \langle X_{ij}^{(a)} X_{ij}^{(b)} \rangle \tag{18}$$

for the centralised variables $X_{ij}^{(a)} \equiv \phi_{ij}^{(a)} - \langle \phi_{ij}^{(a)} \rangle$.

Here, we mostly use temporal PCA which involves time averaging rather than spatial PCA. The choice of averaging depends on the data sets and the requirements of the analysis. Assuming all phases measure the same interface, there is one eigenvector $\mathbf{v}_k$ of $C_{ab}$ with eigenvalue $\lambda_k$ much larger than the others, hereby referred to as the principal component. The confidence of the PCA,

$$sg_k = \frac{|\lambda_k|}{\sum_{a \neq k} |\lambda_a|}, \tag{19}$$

where $\lambda_k$ is the eigenvalue of the principal component, representing how well the height fluctuations can be aligned with the data $\mathbf{p}$. As per the phase equations (6), the principal component $\mathbf{v}_k$ fixes the ratios of the phase fields. Hence, to get all of the phase prefactors, it is required to estimate one of them and locate it in $\mathbf{v}_k$. For example, in the analysis presented in the paper, $RB_1$ is the easiest peak to identify by its brightness with respect to the others. The ratio between the entries of $\mathbf{v}_k$ and the $RB_1$ entry, $\mathbf{v}_{k,0}$, equals the ratio between the prefactors $\alpha_m$, that is, $\frac{\mathbf{v}_{k,m}}{\mathbf{v}_{k,0}} = \frac{\alpha_m}{\alpha_0}$. Hence, it suffices to know the prefactor of one of the peaks, in the latter case $\alpha_{RB_1} \equiv \alpha_0$, to estimate the

remaining prefactors $\alpha_{AB} \equiv \alpha_m$.



\end{document}